\documentclass{moriond}

\usepackage{comment}

\def\be{\begin{equation}}
\def\ee{\end{equation}}
\def\bea{\begin{eqnarray}}
\def\eea{\end{eqnarray}}

\newcommand{\Photo}{}

\begin{document}
\vspace*{4cm}
\title{Lepton-Pair Scattering With an Off-Shell and an On-Shell Photon at Two Loops in Massless QED}

\author{Simone Zoia}

\address{CERN, Theoretical Physics Department, CH-1211 Geneva 23, Switzerland}

\maketitle\abstracts{
I present the computation of the two-loop amplitudes for the scattering of a lepton pair with an off-shell and an on-shell photon in massless QED.\cite{Badger:2023xtl}
We apply modern techniques developed to tackle QCD amplitudes with many scales:
we express the Feynman integrals in terms of a basis of special functions, and reconstruct the amplitudes from numerical finite-field evaluations.
Our results complete the amplitude-level ingredients for the N3LO predictions of electron-muon scattering needed to meet the precision target of the future MUonE experiment.
}

\par\textsc{CERN-TH-2024-059}

\section{Motivation: the Muon Anomalous Magnetic Moment}
\label{sec:Introduction}

\begin{figure}[t]
\centering
\begin{minipage}{.49\textwidth}
  \centering
  \includegraphics[width=.56\linewidth]{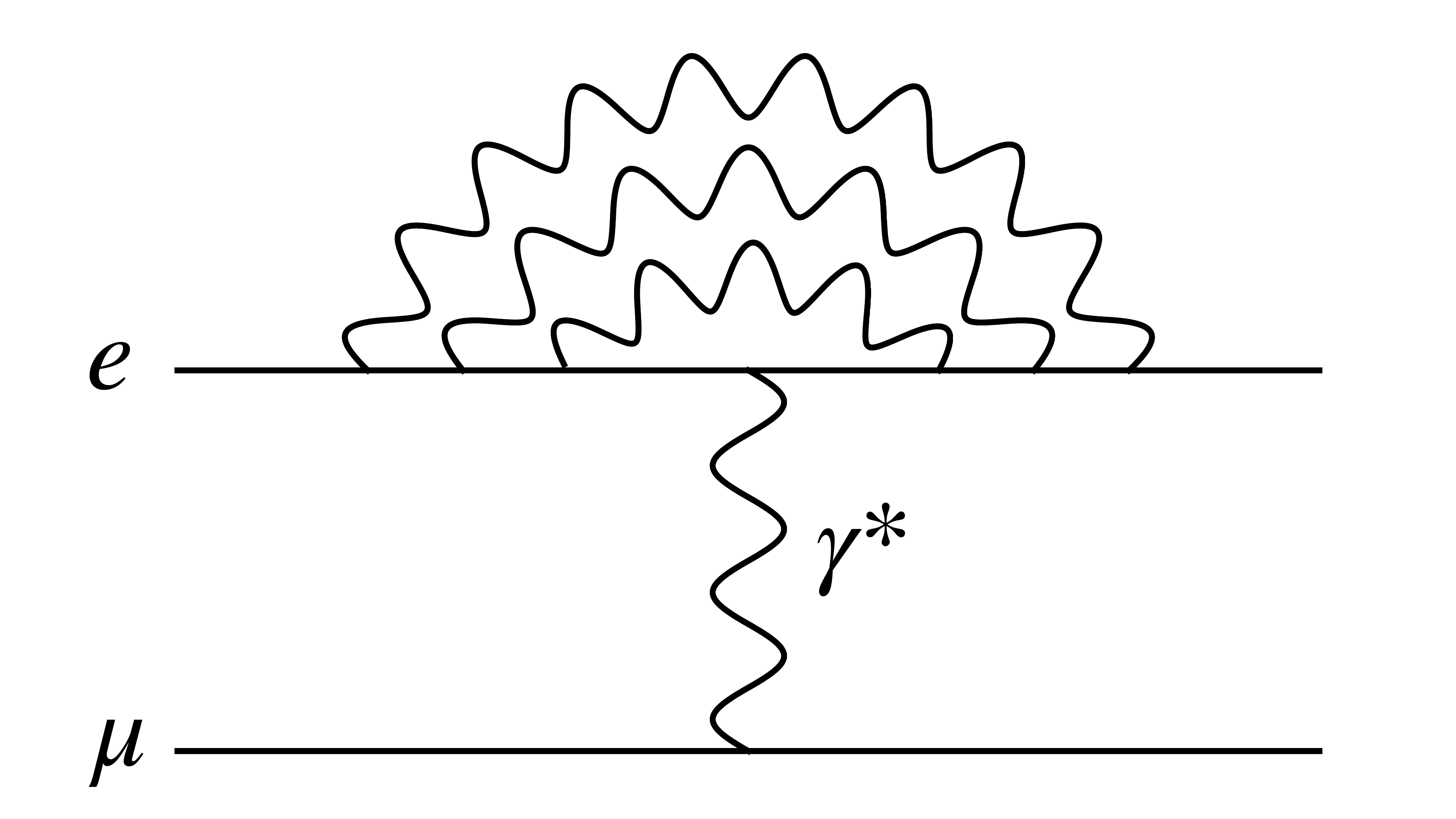}
  \caption[]{sample Feynman diagram contributing to the electron-line corrections to \mbox{$e\mu \to e\mu$} at N3LO in QED.}
   \label{fig:ElectronLine}
\end{minipage}%
\hfill
\begin{minipage}{.47\textwidth}
  \centering
  \includegraphics[width=.46\linewidth]{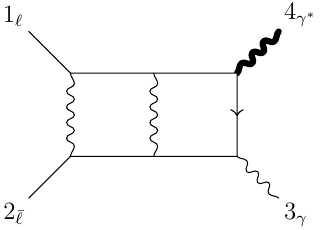}
  \caption[]{sample two-loop Feynman diagram for the process in Eq.~(\ref{eq:scatter}).}
  \label{fig:FeynDiag}
\end{minipage}
\end{figure}

The muon anomalous magnetic moment $a_\mu$ is at the heart of one of the longest-standing tensions among experimental,\cite{Muong-2:2021ojo} Standard Model (SM) data-driven,\cite{Aoyama:2020ynm} and lattice QCD~\cite{Borsanyi:2020mff} results.
The main source of error in SM prediction is the Hadronic Vacuum Polarisation (HVP) contribution, $a^{\rm HVP}_{\mu}$.
It is therefore crucial to obtain independent determinations of $a^{\rm HVP}_{\mu}$.
The planned MUonE experiment~\cite{Abbiendi:2016xup} aims to measure the hadronic running of the electromagnetic coupling using elastic electron-muon scattering ($e\mu \to e\mu$), which will enable a new and precise determination of $a_{\mu}^{\rm HVP}$.\cite{CarloniCalame:2015obs}
Full NNLO QED predictions for $e\mu \to e\mu$ have been completed recently,\cite{Broggio:2022htr} highlighting the need for N3LO corrections to achieve MUonE's precision goal. 
The dominant contribution comes from the electron-line corrections, i.e., corrections to the sub-process with the muon line stripped off ($e\to e \gamma^*$, see Fig.~\ref{fig:ElectronLine}).
The main missing ingredient is the RVV (real double-virtual) matrix element ($e\to e \gamma \gamma^*$ at two loops).\cite{Fael:2023zqr}

The two-loop amplitudes for $e\to e \gamma \gamma^*$ play a role also to compute $a^{\rm HVP}_{\mu}$ from the $e^+ e^- \to \gamma^* \to {\rm hadrons}$ data.
In the energy scan measurements, these amplitudes are part of the RVV corrections at N3LO.
In the radiative return measurements, instead, we measure a photon in addition to the hadrons, and thus they contribute to the VV corrections at NNLO.
In these cases, however, the bottleneck is in the hadronic~decay.

The two-loop amplitudes for $e\to e \gamma \gamma^*$ in massless QED are fairly simple by today's state of the art, and could be extracted from known results.\cite{Garland:2001tf}
We decided to perform a direct computation making use of the modern analytic techniques developed to tackle processes with many scales in QCD (see S.~Badger's talk about ${\rm t\bar{t}} + {\rm jet}$ production\cite{Badger:2024fgb}).
This allowed us to obtain compact analytic expressions which can be evaluated numerically efficiently, and gave us an opportunity to spread this technology to the community working on QED/EW amplitudes.
As a proof of the relevance of this work, another direct computation followed shortly after ours.\cite{Fadin:2023phc}

\section{Tackling Algebraic and Analytic Complexity in the Scattering Amplitudes}
\label{sec:Workflow}

\begin{figure}[t]
\begin{minipage}{0.32\linewidth}
\centerline{\includegraphics[width=0.6\linewidth]{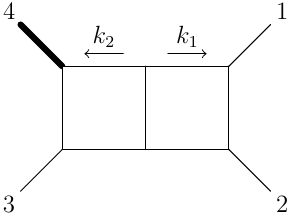}}
\end{minipage}
\hfill
\begin{minipage}{0.32\linewidth}
\centerline{\includegraphics[width=0.6\linewidth]{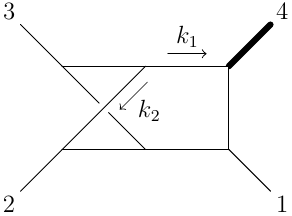}}
\end{minipage}
\hfill
\begin{minipage}{0.32\linewidth}
\centerline{\includegraphics[width=0.6\linewidth]{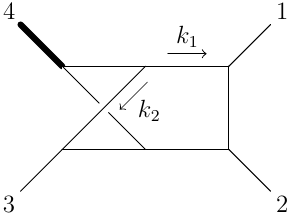}}
\end{minipage}
\caption[]{graphs representing the relevant two-loop Feynman integral families.}
\label{fig:IntegralFamilies}
\end{figure}

We compute the two-loop helicity amplitudes $A^{(2)}$ for the process
\begin{equation}
    \label{eq:scatter}
    0 \to \ell(p_1) + \bar{\ell}(p_2) + \gamma(p_3) + \gamma^*(p_4) \,,
\end{equation}
where $\ell$ is an on-shell massless lepton and $\gamma$ ($\gamma^*$) an on-shell (off-shell) photon.
The external momenta $p_i$ satisfy the on-shell conditions $p_i^2 = 0$ for $i=1,2,3$.
We choose the independent invariants as
$\vec{s} = \bigl( s_{12}, s_{23}, s_4 \bigr)$,
with $s_{i\ldots j} = (p_i+\ldots+p_j)^2$.
We use dimensional regularisation with $D=4-2\epsilon$ spacetime dimensions.
The starting point are the Feynman diagrams, which we generate with \texttt{QGRAF}~\cite{Nogueira:1991ex} (see Fig.~\ref{fig:FeynDiag} for an example).
We then manipulate them to write each helicity amplitude as a linear combination of scalar Feynman integrals of the families shown in Fig.~\ref{fig:IntegralFamilies}.\footnote{The other needed two-loop families are expressed in terms of those in Fig.~\ref{fig:IntegralFamilies} and products of one-loop integrals.}
Their coefficients are rational functions of $\vec{s}$ and $\epsilon$.
By solving the integration-by-parts (IBP) relations,\cite{Chetyrkin:1981qh,Laporta:2001dd} we express the scalar integrals in terms of fewer, linearly independent `master' integrals (MIs).
We generate the IBP relations using \texttt{LiteRed},\cite{Lee:2012cn} and solve them with \texttt{FiniteFlow}.\cite{Peraro:2019svx}
Finally, we expand around $\epsilon=0$ up to order $\epsilon^0$, obtaining
\begin{equation} \label{eq:Aexpr}
A^{(2)}(\vec{s}; \epsilon) = \sum_{k=-4}^0 \sum_i \epsilon^k \, r_{ki}(\vec{s}) \, F_i(\vec{s}) \,,
\end{equation}
where $r_{ki}$ ($F_i$) are rational (special) functions. 
This separation reflects the two kinds of complexity which plague this type of analytic computation: the algebraic and the analytic complexity. 

\medskip

While the rational functions are simple from the analytic point of view, their sheer size ---~the \emph{algebraic complexity}~--- can make them difficult to handle.
This is particularly problematic in the intermediate stages:
while the input (the Feynman diagrams) and the final result (the amplitude) are comparatively compact, the intermediate expressions may swell to the point of jeopardising the computation.
Two simple yet important insights allow us to sidestep this bottleneck. 
First, we are not interested in knowing the intermediate results analytically.
Second, the expression swell affects only the symbolic computations, and can be deflated by instead evaluating the rational functions numerically.
Leveraging these ideas, Peraro~\cite{Peraro:2016wsq} pioneered a new approach, which replaces the symbolic manipulations of rational functions with numerical evaluations over finite fields, i.e., integers modulo a prime number.
Modular arithmetic allows us to avoid both the intrinsic loss of accuracy of floating-point numbers, and the computationally expensive arbitrary-precision arithmetic of exact rational numbers.
The analytic expression of the final result is then obtained from sufficiently many numerical evaluations by means of functional reconstruction algorithms.
We perform all the rational operations on rational functions numerically over finite fields, and only reconstruct the rational coefficients of the final result in Eq.~(\ref{eq:Aexpr}), thus sidestepping the intermediate expression swell.
The entire workflow is implemented within \textsc{FiniteFlow}, adopting a number of optimisations to simplify the functional reconstruction.\cite{Zoia:2023nup}

\medskip

The special functions arise from the loop integrations, and are instead characterised by their \emph{analytic complexity}.
Here, the difficulty is also conceptual other than computational. 
Even just understanding which class of special functions appears in a given amplitude may be challenging, let alone evaluating and manipulating them.
Moreover, special functions satisfy functional relations. A toy example is $\log (x y) = \log x + \log y$ ($x,y>0$) for the logarithm.
A representation of an amplitude in terms of special functions may thus be redundant.
This leads to a more complicated expression and to an unstable evaluation, as the cancellations only occur numerically.
In the most complicated cases, we lack the mathematical technology to overcome these issues.
In the best understood cases, the most successful approach is the method of differential equations (DEs)~\cite{KOTIKOV1991158,Gehrmann:1999as,Bern:1993kr} in the canonical form.\cite{Henn:2013pwa}
The idea is to choose the MIs $\vec{g}$ such that they satisfy a system of partial DEs of the~form
\begin{equation} \label{eq:canDEs}
\mathrm{d}\vec{g}\left(\vec{s}; \epsilon\right) = \epsilon \, \sum_{i} a_i \, \mathrm{d} \log\left(W_i\left(\vec{s}\right)\right) \cdot \vec{g}\left(\vec{s};\epsilon\right) \,,
\end{equation}
where the $a_i$'s are constant rational matrices, and the \emph{letters} $W_i(\vec{s})$ are algebraic functions of the invariants $\vec{s}$ (e.g., $s_{12}$, $s_{23}$, $s_{12}+s_{23}$ etc.).
Finding such MIs is in general difficult, but the integral families in Fig.~\ref{fig:IntegralFamilies} are simple by today's standards, and indeed they had been computed already long ago.\cite{gehrmann:2000zt,gehrmann:2001ck}
We instead focus on how to solve Eq.~(\ref{eq:canDEs}) so as to overcome the issues above.
Also in this case, we follow an approach which has proven successful in the context of multi-scale QCD amplitudes.\cite{Chicherin:2020oor,Chicherin:2021dyp,Abreu:2023rco}
By making use of a mathematical technique called Chen iterated integral, we construct a basis of \emph{algebraically independent} and \emph{irreducible} special functions in which all MIs (including all their required crossings) can be expressed.
This guarantees that we have a unique and compact representation of the amplitudes, and that we evaluate the smallest possible number of special functions. 
In order to evaluate them numerically, we express the basis functions in terms of multiple polylogarithms (MPLs).
We then apply a mathematical technique called Lyndon decomposition to further optimise the expressions.
In summary, a high degree of optimisation is unlocked by a deep mathematical understanding of the relevant special~functions.

\section{Conclusions and Outlook}
\label{sec:Conclusions}

We computed the two-loop amplitudes for $0\to\ell\bar\ell\gamma\gamma^*$ in massless QED in terms of a basis of MPLs that are suitable for efficient evaluation. 
We used finite-field techniques to sidestep the intermediate expression swell.
Our results are ready for deployment in phenomenology, and indeed have already been implemented in \texttt{McMule}~\cite{Banerjee:2020rww} to provide the RVV (electron-line) corrections to $e \mu \to e \mu$.
These results pave the way to N3LO predictions for the future MUonE~experiment.

\section*{Acknowledgments}

I am grateful to Simon Badger, Jakub Kry\'s and Ryan Moodie for collaboration on this project.
This project received funding from the European Research Council (ERC) under the European Union's Horizon 2020 research and innovation programme (grant agreement No.~772099), and Horizon research and innovation programme (grant agreement No.~101105486).

\end{document}